\documentclass[prb,twocolumn,showpacs,floatfix,preprintnumbers,amsmath,amssymb,superscriptaddress]{revtex4-1}
\usepackage{graphicx}
\usepackage{dcolumn}
\usepackage{bm}
\usepackage{color}
\usepackage[latin1]{inputenc}
\usepackage[urlcolor=blue]{hyperref}
\hypersetup{backref, colorlinks=true}

\usepackage{color}
\usepackage[latin1]{inputenc}

\newcommand{\EIS}{EuIr$_2$Si$_2$}
\newcommand{\ENP}{EuNi$_2$P$_2$}

\begin{document}


\title{Valence effect on the thermopower of Eu systems}

\author{U.~Stockert}\email{ulrike.stockert@cpfs.mpg.de}\affiliation{Max Planck Institute for Chemical Physics of Solids, N\"{o}thnitzer Straße 40, D­-01187 Dresden, Germany}
\author{S.~Seiro}\affiliation{Max Planck Institute for Chemical Physics of Solids, N\"{o}thnitzer Straße 40, D­-01187 Dresden, Germany}\affiliation{Leibniz IFW Dresden, Helmholtzstr. 20, D-01069 Dresden, Germany}
\author{N.~Caroca-Canales}\affiliation{Max Planck Institute for Chemical Physics of Solids, N\"{o}thnitzer Straße 40, D­-01187 Dresden, Germany}
\author{E.~Hassinger}\affiliation{Max Planck Institute for Chemical Physics of Solids, N\"{o}thnitzer Straße 40, D­-01187 Dresden, Germany}\affiliation{TU M\"{u}nchen, James-Franck-Str. 1, 85748 Garching, Germany}
\author{C.~Geibel}\affiliation{Max Planck Institute for Chemical Physics of Solids, N\"{o}thnitzer Straße 40, D­-01187 Dresden, Germany}

\date{\today}

\begin{abstract}
We investigated the thermoelectric transport properties of \ENP~and \EIS~in order to evaluate the relevance of Kondo interaction and valence fluctuations in these materials. While the thermal conductivities behave conventionally, the thermopower curves exhibit large values with pronounced maxima as typically observed in Ce- and Yb-based heavy-fermion materials. However, neither the positions of these maxima nor the absolute thermopower values at low temperature are in line with the heavy-fermion scenario and the moderately enhanced effective charge carrier masses. Instead, we may relate the thermopower in our materials to the temperature-dependent Eu valence by taking into account changes in the chemical potential. Our analysis confirms that valence fluctuations play an important role in \ENP~and \EIS. 
\end{abstract}


\maketitle

\subsection{Introduction}

Intermetallic Eu compounds exhibit various exotic phenomena such as non-integer valence, valence transitions and valence fluctuations. They are related to the existence of two Eu configurations with relatively small energetic distance: the nonmagnetic one of Eu$^{3+}$ ( $4f^6$) and the magnetic one of Eu$^{2+}$ ($4f^7$).~\cite{Onu17} A non-integer average Eu valence may arise due to several physical reasons: (1) A trivial case are mixed-valent systems containing more than one crystallographic Eu site with different valence as, for example, Eu$_3$O$_4$.~\cite{Rau66} More interesting are materials with a single Eu site: (2) Valence-fluctuating (VF) systems such as Eu$_4$Pd$_{29+x}$B$_8$~\cite{TB-16-1} exhibit a non-integer valence due to thermal fluctuations between the two close-lying integer Eu configurations. The respective characteristic energy scale is given by the valence fluctuation temperature $T_\mathrm{VF}$. (3) In intermediate valent (IV) systems a non-integer Eu valence originates from hybridization between the Eu $4f$ and the conduction electron states due to Kondo interaction, for example in EuCu$_2$(Ge$_{1-x}$Si$_x$)$_2$ close to $x = 0.7$.~\cite{TB-04-6,B-18-3} In this case the characteristic energy scale corresponds to the Kondo temperature $T_\mathrm{K}$. Unfortunately, the notation in literature is not consistent, and frequently the terms "valence fluctuating", "intermediate valent", and sometimes even "mixed valent" are used as synonyms. The situation is also complicated by the fact that more than one effect may be relevant in a material. For example, the two distinct Eu configurations of a VF material usually experience also some energetic broadening due to interactions with the conduction electron states. Another possibility are mixed-valent systems with valence fluctuations, which may even be site-dependent as discussed for Eu$_3$Pd$_{20}$Ge$_6$.~\cite{B-09-1}

\ENP~and \EIS~have been classified as systems with VF or IV character. They exhibit a strongly temperature-dependent Eu valence $\nu$ that remains fractional down to lowest temperature $T$ as revealed by M\"{o}ssbauer spectroscopy.~\cite{ENP-85-1, ENP-86-2} Both materials have moderately enhanced effective charge carrier masses indicative of significant hybridization between Eu $4f$ and conduction electron states.~\cite{ENP-11-1, ENP-95-1} Recently, \ENP~and \EIS~attracted considerable interest due to the question of how this hybridization, which is usually observed in Ce and Yb-based Kondo systems, is related to the fluctuating Eu valence.~\cite{ENP-19-2,ENP-18-1,ENP-19-1,ENP-13-1}

In fact, some of the low-temperature properties of \ENP~and \EIS~can be explained in the framework of a simple model for VF materials: the interconfiguration fluctuation model.~\cite{B-75-1, B-80-2} In its simplest form it considers Eu$^{2+}$ and Eu$^{3+}$  states that are close in energy. A phenomenological spin-fluctuation energy is introduced to account also for interactions with conduction electron states. The model yields a temperature-dependent effective Eu valence via thermal excitations from the low-energy Eu$^{3+}$ into the higher Eu$^{2+}$ state. It has been applied to fit the magnetic susceptibility of \ENP. \cite{ENP-85-1} Likewise, the $T$-dependences of the magnetic susceptibility and valence of \EIS~can be described within this model, although not with a single set of parameters.~\cite{ENP-19-2}

Despite this success of the interconfiguration fluctuation model, there are several observations which require a more complex picture. Recent NMR measurements on \ENP~revealed a strong temperature dependence of the spin-lattice relaxation rate $1/T_1$ that cannot be explained assuming a simple combination of Eu$^{2+}$ and Eu$^{3+}$ states.~\cite{ENP-18-1} The enhanced effective charge carrier masses of \ENP~and \EIS~mentioned above suggest a significant hybridization between $4f$ and conduction electron states at low $T$ due to Kondo interaction. In fact, the formation of heavy bands resulting from a hybridization between Eu $4f$ and Ni $3d$ states has been observed directly in photoemission spectroscopy on \ENP.~\cite{ENP-09-1} Indication for a hybridization of $4f$ and conduction electron states has been obtained also from optical conductivity measurements on \ENP~and \EIS.~\cite{ENP-12-1} In addition, electrical resistivity and Hall effect measurements on \ENP~have been interpreted with the formation of a heavy-fermion (HF) state as typically observed in Ce and Yb-based Kondo systems.~\cite{ENP-13-1,ENP-19-1} 

The thermopower is known as a sensitive probe for Kondo scattering in HF systems, where it reaches large absolute values around the characteristic Kondo temperature $T_\mathrm{K}$ and enhanced values of $S/T$ in the zero-temperature limit. Here, we present thermal transport data on \ENP~and \EIS~with a focus on the thermopower $S(T)$ in order to evaluate the relevance of the Kondo interaction and valence fluctuations in these materials. In both compounds we observe indeed large thermopower values, larger than in other VF Eu compounds. However, the temperature dependencies $S(T)$ cannot be explained by simple models for HF systems with Kondo interaction. Instead, we may relate $S(T)$ directly to the temperature dependence of the Eu valence by considering changes in the chemical potential. These results suggest that \ENP~and \EIS~cannot be understood within a pure Kondo scenario for HF compounds, but that thermal valence fluctuations between different Eu configurations have to be taken into account and are the main origin of the large thermopower in these materials.


\subsection{Experimental details}

We investigated single crystals of \ENP~and \EIS~grown by a flux method as described in Refs.~\onlinecite{ENP-11-1},\onlinecite{ENP-09-1}. Thermal conductivity $\kappa$, thermopower $S$, and electrical resistivity $\rho$ were measured simultaneously using the thermal transport option (TTO) of a commercial physical property measurement system (PPMS from Quantum Design). The electrical and heat currents were applied within the $ab$ plane of our plate-like single crystals. Measurements have been performed in the temperature range from 2~K to 300~K. With increasing temperature the heat loss via radiation becomes large. This leads to a strong upturn of our thermal conductivity data above 200~K. Therefore, we do not show thermal conductivity data in this temperature range. The effect is not relevant for thermopower measurements, which allows us to discuss $S(T)$ up to 300~K.

The geometry factor of our samples has a large uncertainty due to the small crystal size of about $4 \times 1 \times 0.1$ mm$^3$ and the relatively large contacts with a width of approximately 0.5 mm required for a good thermalization of the thermometers. Therefore, we scaled our electrical resistivity data to results from AC transport measurements on crystals from the same batches with an optimized geometry. We used the same scaling factor for our thermal conductivity data. The analysis of $\rho$ and $\kappa$ measured on the same contacts reduces uncertainties of the geometry factor in the estimation of the electronic thermal conductivity from the Wiedemann-Franz law.


\subsection{Results}

\begin{figure}[tb]
	\begin{center}
		\includegraphics[width=0.48\textwidth]{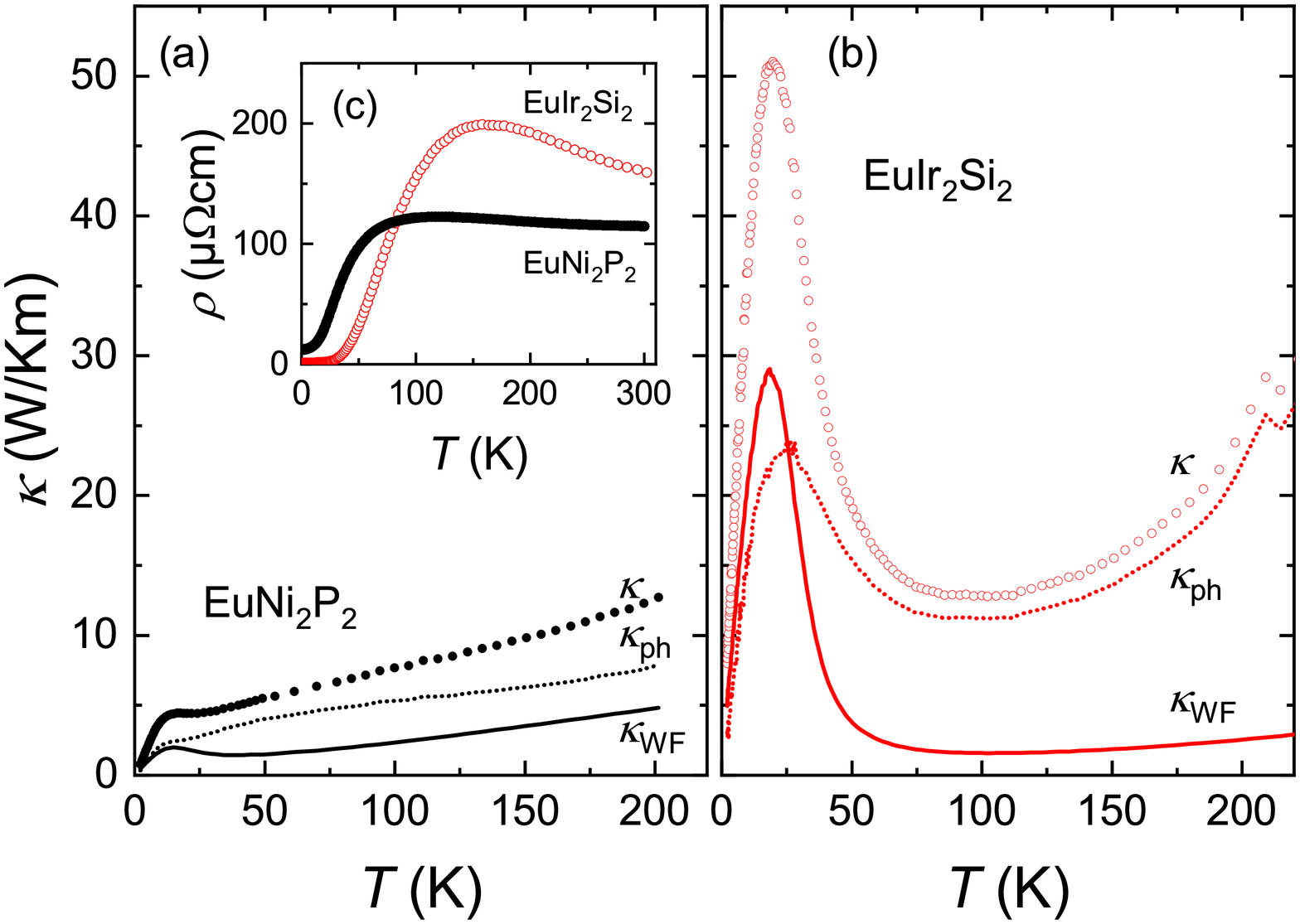}
	\end{center}
	\caption{(Color online) Temperature dependencies of the total thermal conductivity $\kappa$ and the estimated electron ($\kappa_{\mathrm{WF}}$) and phonon contributions ($\kappa_{\mathrm{ph}}$) for \ENP~(a) and \EIS~(b). The inset (c) show the corresponding electrical resistivities measured simultaneously on the same contacts. \label{KvsT}}
\end{figure}

The electrical resistivities $\rho(T)$ of \ENP~and \EIS~are plotted in the inset (c) of Fig.~\ref{KvsT}. Our results are similar to data reported in literature.~\cite{ENP-13-1,ENP-19-2} 
Below room temperature we observe a negative $\partial \rho / \partial T$ for both materials. Maxima are reached at about 120~K (\ENP) and 160~K (\EIS). Towards lower $T$ the electrical resistivities strongly decrease. The residual resistivity ratio RRR determined as $\rho (300K)/ \rho (2K)$ is significantly larger for the \EIS~sample (160) than for the \ENP~sample (9.3).

The thermal conductivities $\kappa (T)$ of \ENP~and \EIS~are plotted in Fig.~\ref{KvsT}a and b, respectively. \EIS~has a significantly larger thermal conductivity than \ENP. It exhibits a pronounced maximum around 20~K, while only a small hump is seen in this temperature region for \ENP. In order to evaluate the origin of this difference we estimated the thermal conductivity contributions from charge carriers ($\kappa_{\mathrm{el}}$) and phonons ($\kappa_{\mathrm{ph}}$) using the Wiedemann-Franz (WF) law. We assume that $\kappa_{\mathrm{el}} = \kappa_{\mathrm{WF}} = L_0 T/\rho$ with the Lorenz constant $L_0$ and determine $\kappa_{\mathrm{ph}}$ as $\kappa_{\mathrm{ph}} = \kappa - \kappa_{\mathrm{WF}}$. The results are also shown in Fig.~\ref{KvsT}. We find that the small bump in the thermal conductivity of \ENP~can be attributed almost completely to $\kappa_{\mathrm{el}}$, while $\kappa_{\mathrm{ph}}$ exhibits only a weak temperature dependence. By contrast, the large maximum in the thermal conductivity of \EIS~is most probably due to maxima of both contributions, $\kappa_{\mathrm{el}}$ and $\kappa_{\mathrm{ph}}$. Since the WF law is strictly valid only in the zero-$T$ limit we cannot exclude that $\kappa_{\mathrm{el}}$ of \EIS~is larger than our estimate. However, it seems unlikely that $\kappa_{\mathrm{el}}$ may account for the full thermal conductivity maximum. Most probably, the phonon thermal conductivity of \EIS~exhibits a maximum as expected for clean single crystals. The absence of such a maximum for \ENP~is attributed to a lower sample quality, where scattering from defects leads to a
suppression of $\kappa_{\mathrm{ph}}$. This is also in line with the significantly smaller RRR for our \ENP~sample compared to the \EIS~crystal. 

\begin{figure}[tb]
	\begin{center}
		\includegraphics[width=0.48\textwidth]{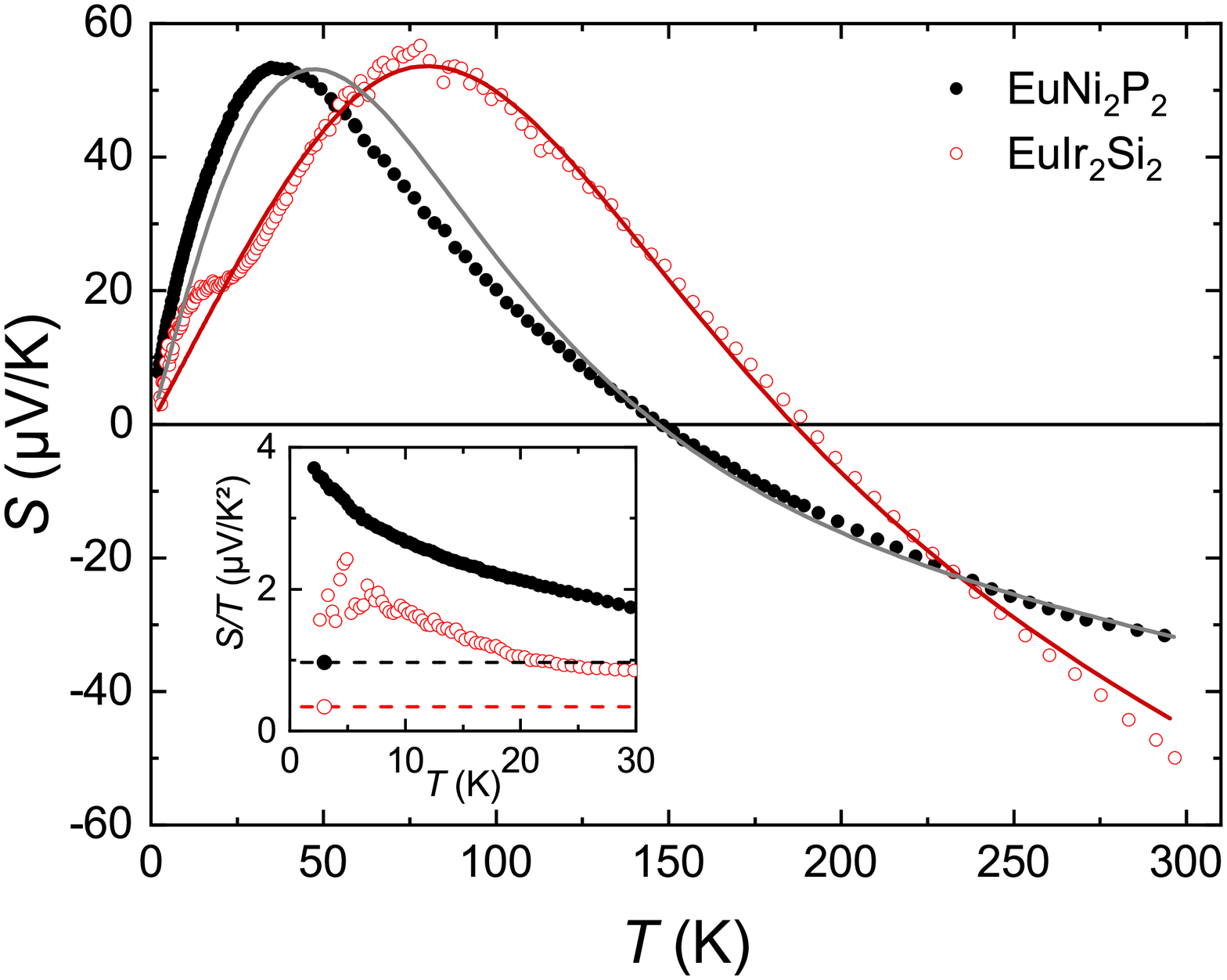}
	\end{center}
	\caption{(Color online) Temperature dependence of the thermopower $S(T)$ for \ENP~and \EIS. The lines in the main plot are fits to the data taking into account the temperature dependence of the valence as explained in the text. The hump at around 20~K for \EIS, which is not reproduced by the fitted curve, is ascribed to phonon drag. The inset shows the low-$T$ part as $S/T$ (symbols) in comparison to the zero-temperature limit of $S_{\mathrm{d}}/T$ (lines) calculated from the effective charge carrier masses as explained in the text. \label{SvsT}}
\end{figure}

The thermopower $S(T)$ of \ENP~and \EIS~is plotted in Fig.~\ref{SvsT} for the temperature range between 2~K and 300~K. Both compounds exhibit a similar qualitative temperature dependence of $S(T)$ with large, positive values at low $T$ and a change to negative values at higher $T$. Maxima are observed at $T_\mathrm{max} = 37$~K and 81~K for \ENP~and \EIS, respectively. The thermopower of \EIS~exhibits an additional hump at about 20~K, i.e., close to the position of the maximum in our estimate of $\kappa_{\mathrm{ph}}$. Therefore, we attribute this hump to a phonon-drag contribution. No such contribution is expected for \ENP~due to the much lower phonon thermal conductivity in the respective $T$ range. 

At first glance, the thermopower of \ENP~and \EIS~resembles those of other Eu systems discussed either as VF or IV materials (apart from the presumed phonon-drag contribution for \EIS). Eu-based VF systems usually have a positive thermopower with a single maximum close to the valence fluctuation temperature $T_\mathrm{VF}$~\cite{Jac82} as seen, for instance, in Eu$_3$Pd$_{20}$Ge$_6$ \cite{TB-02-6} and Eu$_4$Pd$_{29+x}$B$_8$.~\cite{TB-16-1} Our thermopower curves for \ENP~and \EIS~exhibit a similar behavior, however, with some important differences: (1) Our absolute thermopower values are much larger exceeding 50~$\mu$V/K at the maxima compared to about 10~$\mu$V/K in Eu$_3$Pd$_{20}$Ge$_6$ and Eu$_4$Pd$_{29+x}$B$_8$. In fact, the maxima of $S(T)$ of \ENP~and \EIS~are comparable to those of Ce- and Yb-based IV systems or of Eu systems discussed as HF compounds with strong Kondo interaction such as EuCu$_2$Si$_2$.~\cite{TB-04-6} (2) Above 150~K our thermopower curves change sign to negative values. (3) Moreover, the maxima in $S(T)$ are observed at rather low $T$ compared to $T_\mathrm{VF}$. An estimate for $T_\mathrm{VF}$ is given by the excitation energy $E_\mathrm{exc}$ between the two Eu configurations of the interconfiguration fluctuation model. Fits to M\"{o}ssbauer spectroscopy and magnetic susceptibility data of \ENP~\cite{ENP-85-1,ENP-16-2} yielded values between 160~K and 190~K for  $E_\mathrm{exc}/k_{\mathrm{B}}$ compared to $T_\mathrm{max} = 37$~K. Respective values for $E_\mathrm{exc}/k_{\mathrm{B}}$ of \EIS~have been determined from susceptibility and X-ray absorption measurements on crystals grown under the same conditions as those used in our study. The results correspond to 269~K and 390~K, whereas $T_\mathrm{max} = 81$~K.~\cite{ENP-19-2} These numbers, as well as other important parameters are summarized in Tab.~\ref{table1}. It turns out that for both materials the maximum in $S(T)$ is observed at significantly lower $T$ than expected from $E_\mathrm{exc}$. Actually, $T_\mathrm{max}$ is much closer to the spin-fluctuation temperature $T_\mathrm{sf}$ determined from the same fits to the interconfiguration-fluctuation model: $T_\mathrm{sf}$ of \ENP~takes values from 53~K to 80~K, results for \EIS~range from 84 K to 101~K. Since the spin-fluctuation temperature of the model accounts for interactions between the Eu states and the conduction electron states, this observation suggests that the Kondo interaction might be responsible for the $T$ dependence of the thermopower of \ENP~and \EIS.
 
Therefore, and in view of the rather large thermopower values we compare our $S(T)$ curves to those of IV systems. IV materials usually exhibit large absolute thermopower values of about $50-100 \mu$V/K with a maximum around the characteristic Kondo temperature $T_\mathrm{K}$.~\cite{TB-05-3} An estimate for the Kondo temperature of \ENP~has been obtained from specific heat and thermal expansion data.~\cite{ENP-13-1} The value of $T_\mathrm{K} \approx 80$~K is about a factor of 2 larger than $T_\mathrm{max}$. No such evaluation has been performed for \EIS. However, the Sommerfeld coefficient $\gamma _0$ of \EIS~is significantly smaller than that of \ENP~(33 mJ/mole K vs.~103 mJ/mole K \cite{ENP-95-1,ENP-19-2}). Using $T_\mathrm{K} \propto 1/\gamma _0$ we estimate a Kondo temperature of about 250 K for \EIS~compared to $T_\mathrm{max} = 81$~K. It turns out that for both materials the maximum in $S(T)$ is observed at significantly lower $T$ than expected from $T_\mathrm{K}$.
 
 Next, we consider the magnitude of our thermopower data. The diffusion thermopower $S_{\mathrm{d}}$ of a Fermi liquid $S_{\mathrm{d}}/T$ is expected to reach a constant value in the zero-temperature limit. Within a free electron model $S_{\mathrm{d}}/T$ is related directly to the Sommerfeld coefficient $\gamma_0$ via a dimensionless constant $q$ as $S_{\mathrm{d}}/T = q (N_A e)^{-1} \gamma_0$ with $N_A e$ being Faraday's number.~\cite{TB-04-2} For a wide range of compounds, including a number of HF materials, $q$ has been demonstrated to be close to $\pm 1$, whereas the sign of $q$ depends on the type of charge carriers. This observation has been substantiated also theoretically for HF materials.~\cite{TB-05-4} In the inset of Fig.~\ref{SvsT} we show the low-temperature part of our thermopower curves as $S/T$ vs.~$T$. The values expected from the Sommerfeld coefficients are indicated by lines. It turns out that over a considerable temperature range the thermopower of our materials is much larger than the one expected from this simple estimation. In case of \EIS~this may be attributed at least to some extent to the phonon-drag contribution. However, the large thermopower values of \ENP~cannot be explained solely by the hybridization effects responsible for the moderately enhanced Sommerfeld coefficients. In fact, $S/T$ of \ENP~is increasing down to 2~K, without indication for a saturation. This is not surprising, since the valence of the system remains temperature-dependent down to low $T$. The situation is less clear for \EIS~due to the relatively large scattering of the thermopower data. However, the phonon-drag contribution for this sample is expected to mask any linear behavior of the diffusion thermopower in this $T$ range. Altogether, neither the shape nor the absolute thermopower values of \ENP~and \EIS~can be understood in terms of pure Kondo interaction.

\section{Discussion}

\begin{figure}[tb]
	\begin{center}
		\includegraphics[width=0.48\textwidth]{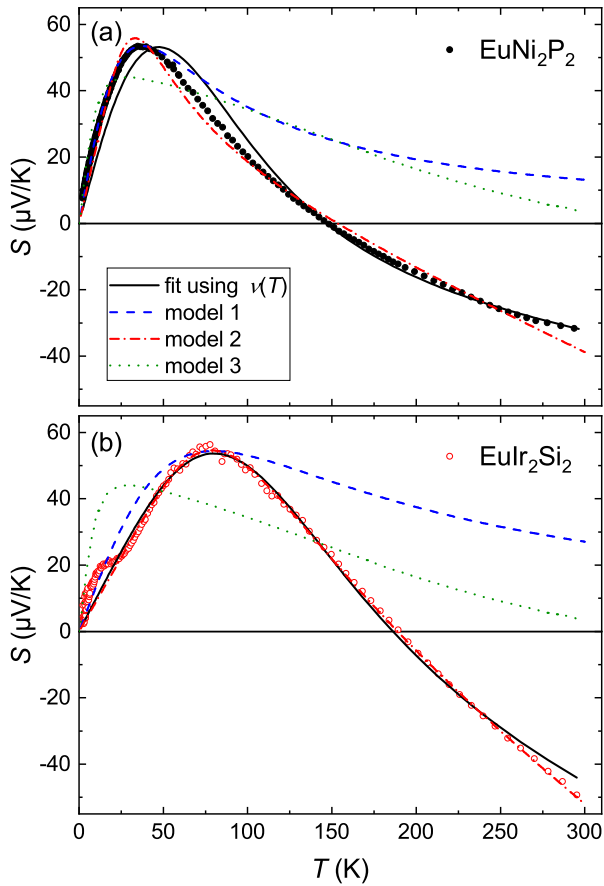}
	\end{center}
	\caption{(Color online) Thermopower of \ENP~(a) and \EIS~(b) in comparison to different calculated or fitted curves. The solid lines correspond to our fits using the temperature dependence of the valence, which is also shown in Fig. 2. The other curves are calculated for different models as explained in the text.\label{SvsTb2}}
\end{figure}

So far, we have discussed the thermopower curves of \ENP~and \EIS~rather qualitatively. In the following we compare our thermopower data to several models considering valence fluctuations and Kondo interaction. Thereafter, we propose a very simple, phenomenological description of our thermopower curves taking into account only the temperature dependence of the valence. Examples of calculated thermopower curves for \ENP~and \EIS~are summarized in 
Fig.~\ref{SvsTb2}.

A very simple expression for the thermopower of $4f$ systems is based on a phenomenological model introduced by Hirst.~\cite{TB-77-1} The model starts with two competing configurations and a finite mixing interaction. The $4f$ density of states (DOS) is approximated by a Lorentzian form with half width $\Gamma$ and the maximum at an energy $\epsilon_0-\epsilon_\mathrm{F}$ with respect to the Fermi level. Within Boltzmann theory and using the relaxation time approximation the thermopower then takes the form $S = AT/(B^2+T^2)$.~\cite{TB-85-3} The parameters $A$ and $B$ are related to $\Gamma$ and $\epsilon_0-\epsilon_\mathrm{F}$. 
Fitting thermopower data of Ce and Yb-based IV materials to this simple relation often yields reasonable agreement with measured curves, especially at intermediate $T$ around the maximum.~\cite{TB-12-1,TB-15-3} However, the model does not describe our thermopower data for \ENP~and \EIS~in a wide temperature range. 
In Fig.~\ref{SvsTb2} we show calculated curves (model 1) for \ENP~and \EIS. $A$ and $B$ have been adjusted to fit the data around the maxima. Although the maxima of $S(T)$ are reproduced quite well, there are significant deviations, especially at higher $T$. In particular, the model cannot account for the sign changes.

Subsequently, the model was extended by including a temperature-dependence of the $4f$ line width $W = 2\Gamma$ of the form $W = T_{\mathrm{sf}} \exp (-T_{\mathrm{sf}}/T)$.~\cite{TB-87-3} The thermopower is then given by $S = c_1TT_0/(T_0^2+W^2)$ with $k_\mathrm{B} T_0 = \epsilon_0-\epsilon_\mathrm{F}$. A second term $c_2T$ is frequently added to account for a band of light charge carriers (d or s type).~\cite{TB-87-3, TB-06-7} However, this approach is problematic since thermopower contributions from different scattering channels do not add directly, but weighted by the electrical resistivities according to the Nordheim-Gorter relation. Moreover, the temperature-dependent $4f$ line width of the model goes to zero in the zero-temperature limit, a situation that is not consistent with the non-integer valence of \ENP~and \EIS. 
Actually, we can describe our thermopower data by this model, however, only when the linear term is large -- much larger than expected for light charge carriers and those typically observed in Ce-based IV systems.~\cite{TB-87-3, TB-06-7} We show the respective curves for \ENP~and \EIS~in Fig.~\ref{SvsTb2} (model 2). For \EIS~the agreement between fit and data is almost perfect. In case of \ENP~the deviations are somewhat larger, and the shape of $S(T)$ is not well reproduced. Since the assumptions of the model are not applicable for our materials, it seems that the good quality of the fit is accidental and probably also facilitated by the relatively large number of -- in total 4 -- free parameters. Other attempts to extend the model, for instance, by introducing also a temperature dependence of $T_0$~\cite{TB-95-4} even further increase the number of fit parameters.

A different starting point to understand the thermopower of IV systems is the Anderson impurity model, which considers the Kondo interaction between conduction and Eu $4f$ electrons. It is frequently used to discuss the thermopower of HF compounds with Ce and Yb,~\cite{TB-05-3} but has been applied also successfully to Eu-based IV materials.~\cite{TB-06-3} The main parameters of this model are the position of the (undisturbed) $4f$ level and the hybridization strength. Thermopower curves were calculated using the non-crossing approximation (NCA) for different sets of parameters. $S(T)$ exhibits a maximum at low temperatures and a sign change at higher $T$ as observed in our measurements. However, these curves are highly assymmetric close to the maxima, which is not in agreement with our experiments. In order to illustrate this we show one curve from Ref.~\onlinecite{TB-06-3} 
in Fig.~\ref{SvsTb2} (model 3). The sign change of this curve is expected above 300~K. 

Another approach to discuss the thermopower of IV Eu systems is based on the Falicov-Kimball model,~\cite{Fal69} which consists of a combination of localized and itinerant states with varying occupation due to thermal fluctuations.~\cite{B-01-4,TB-07-1} However, in the zero-temperature limit this model predicts an integer valence, which is not valid for our materials. In fact, as pointed out in Ref.~\onlinecite{B-01-4} a complete theoretical description of $4f$ fluctuating states requires a combination of the periodic Anderson model with the Falicov-Kimball model. This appears to be rather challenging and has not been accomplished till now.


All models discussed so far are microscopic ones in the sense that they assume either a certain shape of the DOS or a specific type of interaction and then calculate the thermopower. In what follows we use a different approach, namely we evaluate directly the effect of the valence change with temperature on the thermopower without considering the reason for the valence change. In fact, any temperature gradient in our materials is accompanied by a valence gradient because of the strong $T$ dependence of $\nu$. This gives rise to a thermopower contribution $S_{\mathrm{VF}}$ due to the variation of the number of available charge carriers or, more precisely, the change of the chemical potential $\mu$ along the samples. It has been shown,~\cite{TB-85-4, Schnei84} that this anomalous contribution is approximately proportional to the valence change with temperature $d\nu / dT$. In brief, this relation is obtained from the generalized transport equations by using the electrochemical potential $\mu + e\Phi$ instead of the electrical potential $e\Phi$, $e$ being the electron charge. For simplicity we only consider the isotropic situation, where the transport coefficients are scalar. The thermopower then contains 2 contributions:
\begin{equation}
S = L_{12}/L_{11} - \frac{1}{e} \frac{\mathrm{d}\mu}{\mathrm{d}T} = S_{\mathrm{ref}} + S_{\mathrm{VF}}
\label{transport2}
\end{equation}
$L_{11}$ and $L_{12}$ are the transport coefficients. If $\mathrm{d}\mu/\mathrm{d} T$ can be ignored, then $S =S_{\mathrm{ref}} = L_{12}/L_{11}$, which is the well-known relation for $S$ of (physically) homogeneous materials without a temperature-dependent valence. I.e., $S_{\mathrm{ref}}$ corresponds to the thermopower of a suitable reference material. We approximate this contribution by a linear thermopower $S_{\mathrm{ref}} \propto T$, as expected for a simple metal. The second term of Eq.~\ref{transport2} arises due to the temperature-dependence of the chemical potential. 
Assuming that this is dominated by the valence effect, we may replace the temperature derivative of the chemical potential by the one of the valence. In this case, the anomalous thermopower contribution has the form:
\begin{equation}
S_{\mathrm{VF}} = - \frac{1}{e} \frac{\partial \mu}{\partial \nu } \frac{\mathrm{d} \nu}{\mathrm{d}T} = -A_1 \frac{\mathrm{d} \nu}{\mathrm{d}T}
\label{witt}
\end{equation}
In Refs.~\onlinecite{TB-85-4},~\onlinecite{Schnei84} it was suggested that $\partial \mu/\partial \nu$ should be of the order of the Fermi energy $E_\mathrm{F}$ which was then related to the DOS $D$ as $E_\mathrm{F} \propto 1/\mathrm{D}$. However, since in our case the valence change arises due to excitations from the $J = 0$ state of Eu$^{3+}$ to the $J = 7/2$ state of Eu$^{2+}$, the relevant energy scale is rather the excitation energy $E_{\mathrm{exc}}$, so that $eA_1$ should be of similar magnitude as $E_{\mathrm{exc}}$. Above, we have used this energy already as an estimate for the valence fluctuation temperature $T_\mathrm{VF}$ . The respective values determined from fits to the interconfiguration fluctuation model are given in Tab.~\ref{table1}. 

\begin{table}[t]
\begin{center}
\begin{tabular}{ l     c     c }
  \hline
  compound & \ENP  & \EIS \\
  \hline
  literature data &  &  \\
  $T_\mathrm{sf}$ (K) & 53-80 & 84-101 \\
  $\frac{1}{k_{\mathrm{B}}} E_\mathrm{exc}$ (K) & 160-192 & 269-390 \\
   \hline
  measurement results &  & \\
  $T_\mathrm{max}$ (K) & 37 & 81 \\
  \hline
  fit results &  & \\
  $\frac{e}{k_{\mathrm{B}}} A_1$ (K) & 310 & 306 \\
  $A_2$ ($\mu$V$/$K$^2$) & -0.10 & -0.19 \\
\end{tabular}
  \caption{Summary of physical properties and fit results for \ENP~and \EIS. Literature data have been summarized from Refs.~\onlinecite{ENP-85-1},\onlinecite{ENP-86-2},\onlinecite{ENP-19-2},\onlinecite{ENP-87-1},\onlinecite{ENP-16-2}.}
  \label{table1}
\end{center}
\end{table}

In conclusion we get a very simple relation for the thermopower of materials with a strongly temperature-dependent valence: 
\begin{equation}
S =  A_2 T - A_1 \frac{\mathrm{d} \nu}{\mathrm{d}T}
\label{fit}
\end{equation}
At this point we would like to mention that there is an important difference between our linear-in-$T$ term to the thermopower and the linear thermopower contribution often added in literature to account for light charge carriers -- a procedure that may be questioned as discussed above. In our case the simple sum for the two thermopower contributions stems from using the electrochemical potential, which consists of two components, the electrical and the chemical one. I.e., the sum describes the thermopower of a single channel but with a temperature-dependent number of carriers.

We fitted our thermopower curves for \ENP~and \EIS~to Eq.~\ref{fit} omitting the data below 30~K in case of \EIS. In order to reduce the scattering we used modelled curves for $\nu (T)$ from Ref.~\onlinecite{ENP-85-1} (\ENP) and Ref.~\onlinecite{ENP-19-2} (\EIS) instead of measured valence data to calculate the derivatives $\mathrm{d} \nu/\mathrm{d}T$. The results of this procedure are shown in Fig.~\ref{SvsT} in comparison to the data. 
The curves are also plotted in Fig.~\ref{SvsTb2} (fit using $\nu(T)$) to allow for a direct comparison to the other calculated curves. The fit parameters $(e/k_{\mathrm{B}}) A_1$ and $A_2$ are given in Tab.~\ref{table1}. 

It turns out that the simple relation \ref{fit} describes the measured thermopower curves fairly well. The fit parameters $A_1$ and $A_2$ are mainly determined by the height of the maximum in $S(T)$ and the thermopower value at room temperature. By contrast, the position of the maximum $T_\mathrm{max}$ depends only weakly on the fit parameters and mostly on the derivative $\mathrm{d} \nu/\mathrm{d}T$. $T_\mathrm{max}$ is very well reproduced by the fit, especially for \EIS. This provides strong evidence that the maximum in $S(T)$ is indeed caused by the temperature-dependent Eu valence and not by Kondo interaction. Further confirmation for this picture is given by the values for the fit parameter $A_1$. For \EIS, $(e/k_{\mathrm{B}}) A_1$ actually lies within the range expected from $E_\mathrm{exc}/k_{\mathrm{B}}$. The agreement for \ENP~is less perfect, which is discussed below.
	
The quality of the fits is also illustrated by Fig.~\ref{SvsTb2}, which compares the different calculated thermopower curves and fits to the measured data of \ENP~and \EIS. Only model 2 provides a similar good description of the data as Eq.~\ref{fit}, however by using 4 instead of 2 free parameters. For \EIS~the differences between these two curves and the data are small. In case of \ENP~the actual maximum in $S(T)$ is at slightly lower $T$ than predicted by our fit to Eq.~\ref{fit}. However, the overall shape of $S(T)$ is well reproduced by this relation, while model 2 oscillates around the data.

Alltogether, the agreement between fit and data is better for \EIS~than for \ENP. The same holds for the fit parameter $(e/k_{\mathrm{B}}) A_1$ if compared to $E_\mathrm{exc}/k_{\mathrm{B}}$. These deviations may be partially due to the use of fitted data for $\nu(T)$. Any deviation between these curves and the real valence will also lead to systematic errors in the derivatives $\mathrm{d} \nu / \mathrm{d}T$ used in our fitting of $S(T)$. The valence fit for \EIS~has been obtained from valence data measured on samples grown under the same conditions as those investigated here. By contrast, the valence curve for \ENP~stems from a fit to susceptibility data, which is only an indirect measure for $\nu(T)$. The respective fit parameters reproduced the M\"{o}ssbauer isomer shift with limited accuracy, in particular below 40~K, i.e., in the region of our maximum in $S(T)$. Therefore, we may not exclude that the deviations between the peak positions of data and fit for \ENP~are caused by uncertainties in $\nu(T)$ or due to sample dependencies.

Actually, the quality of our fits to $S(T)$ is rather surprising taking into account the approximations used in the derivation of Eq.~\ref{fit}. Our assumption of a linear reference thermopower is a very simple one. Although $S_{\mathrm{d}} \propto T$ in the free-electron approximation at very low and at high temperatures, these two regimes have different slopes and are connected by a crossover region with deviating behavior.~\cite{B-X-5} Moreover, reference materials for \ENP~and \EIS~in the sense of Eq.~\ref{transport2} may still exhibit a hybridization between $4f$ and conduction electrons. This may be the reason for the large values for $A_2$ obtained from our fits. They are significantly larger than those observed in simple metals, which typically reach thermopower values of 2 to 10~$\mu$V/K at 300~K.~\cite{B-X-5} It may also give rise to a non-linear behavior of $S_{\mathrm{ref}}(T)$. Thus, it appears that deviations from $S_{\mathrm{ref}} \propto T$ are rather to be expected than unusual. Another assumption, which may be questioned, is the temperature-independent relation between $\mu$ and $\nu$. It allowed us to use a constant parameter $A \propto E_\mathrm{exc}$ in Eq.~\ref{witt}. However, early M\"{o}ssbauer spectroscopy experiments on \EIS~suggested a temperature-dependent excitation energy $E_\mathrm{exc}.~$ \cite{ENP-86-2, ENP-87-1} It is, therefore, not trivial that we may fit our data using a constant value for $A_1$. Despite these approximations, Eq.~\ref{fit} provides a very good and simple description of the thermopower of \ENP~and \EIS~with only 2 fit parameters. Systematic investigations on a larger number of materials with a strong temperature dependence of the valence are necessary to decide, whether relation~\ref{fit} can be applied more generally. In fact, the 
VF materials Eu$_3$Pd$_{20}$Ge$_6$ and Eu$_4$Pd$_{29+x}$B$_8$ have rather weak temperature dependencies of their valence.~\cite{B-09-1, TB-16-1} Therefore, $S_\mathrm{VF}$ is expected to be small, which is in line with the relatively small thermopowers observed in these materials.


Our analysis of the thermopower of \ENP~and \EIS~revealed that the large values and the characteristic temperature dependencies are not caused predominantly by Kondo interaction between Eu and conduction electron states, but arise mainly from the strong temperature-dependence of the Eu valence. Both the position and the height of the maxima in $S(T)$ are in line with this scenario. This demonstrates that thermal valence fluctuations between distinct Eu states play an important role in these materials at least above 10~K.

\subsection{Summary}
We studied the thermoelectric transport properties of the valence-fluctuating compounds \ENP~and \EIS. The thermal conductivity of these materials shows no unusual behavior. By contrast, the thermopower of \ENP~and \EIS~exhibits large values and a characteristic temperature dependence. We were able to explain this behavior taking into account the temperature dependence of the Eu valence, which leads to an additional thermopower contribution due to variations in the chemical potential. Further experimental and theoretical investigations on the thermopower of materials with a strongly temperature-dependent valence are necessary in order to evaluate the relevance of this effect in general. Our results confirm that valence fluctuations have to be taken into consideration for an understanding of \ENP~and \EIS. 



\bibliographystyle{PRBstyle}

\end{document}